\documentclass[%
 reprint,
 amsmath,amssymb,
 aps,
]{revtex4-1}

\usepackage{graphicx}
\usepackage{dcolumn}
\usepackage{bm}
\usepackage[mathlines]{lineno}

\usepackage[
scale=0.7, marginratio={1:1, 2:3}, ignoreall,
text={7in,10in},centering,
margin=1.5in,
total={6.5in,8.75in}, top=1.2in, left=0.9in, includefoot,
height=10in,a5paper,hmargin={3cm,0.8in},
]{geometry}

\usepackage{geometry}                		
\usepackage{graphicx}				
\usepackage{amssymb}
\usepackage{enumerate}  
 \usepackage{braket}
 \usepackage{mathpazo}
 \usepackage{amsmath}
 \usepackage{graphicx}
 \usepackage{subfigure}
\graphicspath{ {Images_for_Latex/} }

\usepackage{natbib}

\begin{document}


\title{Quantum Metasurfaces}
\author{R. Bekenstein$^{1,2}$, I. Pikovski$^{1,2,3}$, H. Pichler$^{1,2}$, E. Shahmoon$^{2}$, S. F. Yelin$^{2,4}$, M. D. Lukin$^{2}$}
\affiliation{$^1$ ITAMP, Harvard-Smithsonian Center for Astrophysics, Cambridge, MA 02138  \\ $^2$  Physics Department, Harvard University, Cambridge, MA 02138 \\ $^3$ Department of Physics, Stevens Institute of Technology, Hoboken, NJ 07030 \\ $^4$ Department of Physics, University of Connecticut, Storrs, CT 06269 }


\begin{abstract}

\textbf{
Metasurfaces mold the flow of classical light waves by 
engineering sub-wavelength patterns from  dielectric or metallic thin films. We describe and analyze a method  
in which quantum operator-valued reflectivity can be used to control both spatio-temporal and quantum properties of transmitted and reflected light.  Such a quantum metasurface is realized  by preparing and manipulating entangled superposition states of atomically thin reflectors. Specifically, we show that such a system allows for 
massively parallel quantum operations between atoms and photons and for the generation of highly non-classical states of light, including 
photonic GHZ and  cluster states suitable for quantum information processing. We analyze the influence of imperfections and decoherence, as well as specific implementations based on atom arrays excited into Rydberg states.  
Finally, the extension to quantum metamaterials and emulations of quantum gravitational background for light are discussed.}

\end{abstract}
\maketitle
Recently, significant effort has been directed toward structuring classical light waves using spatial light modulators and metasurfaces \cite{bomzon2002space, yu2011light, kildishev2013planar,chen2016review,rubinsztein2016roadmap}. Such structured light has many potential applications, ranging from micro-machining to manipulating particles \cite{baumgartl2008,yao2011orbital,mathis2012micromachining,schley2014loss}. Further, it can affect interactions in nonlinear optical systems \cite{bekenstein2015optical}. While previous efforts focus on
structuring the spatial or temporal properties of the light \cite{weiner2000femtosecond}, we explore whether the quantum properties of light can be efficiently controlled in a similar manner. 
Apart from fundamental significance, such a technique could have important practical applications in fields 
ranging from quantum information processing and communication to metrology.


In this article, we introduce and analyze a method  
in which quantum operator-valued reflectivity can be used to control both spatio-temporal and quantum properties of transmitted and reflected light. Such quantum metasurfaces are realized by preparing and manipulating entangled states of atomically thin reflectors made of atom arrays and scattering light from it. 
This approach enables a cavity-free multi-mode optical scheme for highly parallel quantum operations on multiple photonic qubits.  
In particular, we analyze a protocol for preparing highly entangled photonic states relevant for quantum information and describe a method for quantum control over specific spatial modes of the light.  \newpage

\begin{figure*}
\centering
{\includegraphics[scale = 0.4]{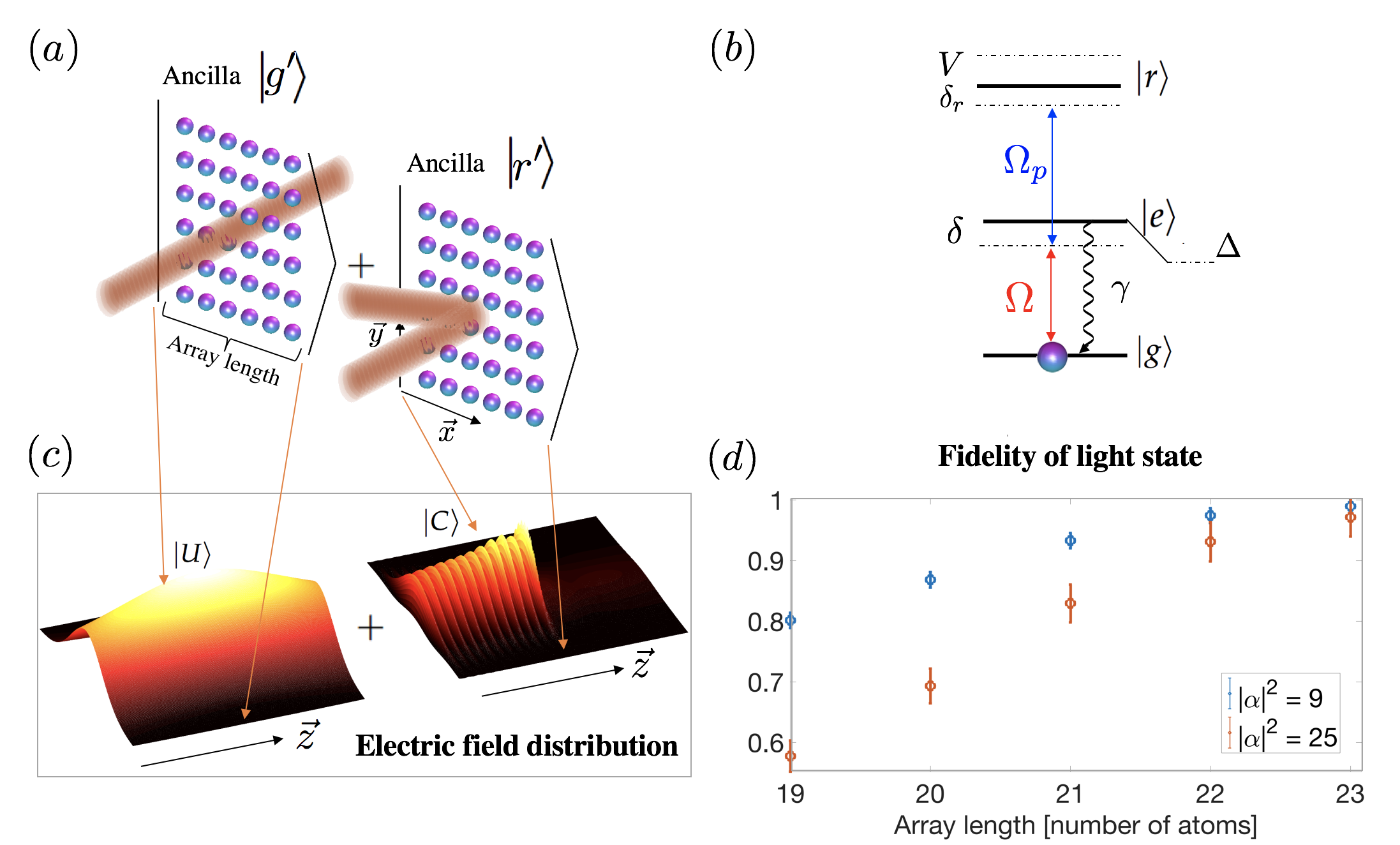}}
\caption{\textbf{Preparing cat states for light with quantum metasurfaces. } (a) Schematic illustration of scattering from the quantum metasurface in a superposition state that perfectly reflects (coupled) and perfectly transmits (uncoupled) the light tuned close to the collective $\ket{g} \rightarrow \ket{e}$ transition (with collective shift $\Delta$), conditioned on the ancilla state. (b) Electronic levels for EIT implementation of a two photon cascade configuration where a control field tuned to the $\ket{e} \rightarrow \ket{r}$ transition is applied, resulting in a transparent array to light tuned to the collective $\ket{g} \rightarrow \ket{e}$ transition. If $\ket{r}$ is shifted due to the interaction with an ancilla, the EIT condition is not fulfilled and the quantum metasurface is reflective. (c) Numerical calculation of the electric field distribution after scattering from the quantum metasurface in the coupled (right) and uncoupled (left) state. (d) Effect of finite array: fidelity of light cat state calculated after the projective measurement of the quantum metasurface state (see SI), as a function of the atom array length (for spacing of $0.2\lambda$ and gaussian beam waist of $1.56\lambda$).}
\end{figure*} 

\noindent
{\bf Key idea \& proposed realization}  

The key idea of this work can be understood by considering a two dimensional array of $N$ atoms placed with sub-wavelength spacing in the $xy$ plane (Fig. 1). 
Such an array can act as an atomically thin mirror for the incident light tuned close to the $\ket{g} \rightarrow \ket{e}$ atomic transition. Near-perfect reflectivity originates from cooperative resonances arising from the dipole-dipole interaction between atoms resulting in collective scattering by the array 
\cite{bettles2016enhanced,shahmoon2017cooperative}. Note that the scattering pattern of the weak incoming light can be effectively engineered via controlled variation of the atomic positions \cite{shahmoon2017cooperative}, in direct analogy to classical metasurfaces.  

To realize a {\it quantum metasurface}, we consider an array prepared in superposition of states with different reflectivity. In such a case, the conditional reflection directly results in  atom-photon entanglement and allows for generating and controlling quantum states of light. To illustrate the main idea, we introduce collective atomic states which are either reflective, since they are coupled ($\ket{C}$) to the light tuned to the collective $\ket{g} \rightarrow \ket{e}$ atomic transition, or transparent, since they are uncoupled ($\ket{U}$) to the latter. We consider an atomic array prepared in the quantum state $|\psi_{QMS} \rangle=\frac{1}{\sqrt{2}}(\ket{U}+\ket{C})$, a superposition of a reflective or transparent state. If a right propagating coherent light wave is launched perpendicular to the array: $\ket{\psi_i} = \ket{\alpha,0}$, where the first mode is right propagating and the second mode is left propagating, then the scattering results in the final state: 

\begin{align}\label{cat}
\ket{\Psi}  = \frac{1}{\sqrt{2}} (\ket{U} \otimes \ket{\alpha,0}+\ket{C} \otimes \ket{t \alpha, r \alpha}),
\end{align} 
where 
$r$ and $t$ are the reflection and transmission coefficients identified with the linear response of the atom array. In particular, when the light is tuned to the cooperative resonance of an infinite array, the perfect conditional reflectivity yields the entangled state: 	 
$\ket{\Psi}  = \frac{1}{\sqrt{2}} (\ket{U} \otimes \ket{\alpha,0}+\ket{C} \otimes \ket{0, -\alpha})$ (see Fig. 1). 
Further control can be obtained by measuring the state of the array. Specifically, a projective measurement in the basis of entangled states $\frac{1}{\sqrt{2}} (\ket{U}\pm\ket{C}$ projects the conditionally scattered light into odd and even cat states for the light $\ket{\psi_f} = \frac{1}{\sqrt{2}} (\ket{\alpha,0} \pm \ket{0,-\alpha})$. 

The key element of this scheme involves preparation \cite{PhysRevLett.122.093601}, manipulation, and measurement of quantum superposition states of the metasurface. These can be implemented by using atom excitation into the Rydberg states. As a specific example, we consider atom array excitation using a two-photon cascade configuration (see Fig. 1b): from a ground state $\ket{g}$ to a Rydberg state $\ket{r}$ via an excited state $\ket{e}$. When a weak (quantum) probe field and a strong (classical) control  field are resonant with respective $\ket{g} \rightarrow \ket{e}$ and $\ket{e} \rightarrow \ket{r}$ transitions, the array is transparent due to electromagnetically induced transparency (EIT), realizing $\ket{U}$. In order to control its reflectivity, a proximal  ancillary, individually controlled atom can be used: if such an atom is prepared in its ground state $\ket{g'}$, it does not affect the array. However if this atom is prepared in a Rydberg state $\ket{r'}$, due to the Rydberg blockade effect it can eliminate EIT for the array atoms and make it highly reflective by shifting the $\ket{r}$ level out of resonance required for the EIT condition (see Fig. 1b). In such a case the array is in its reflective state ($\ket{C}$) (see Fig. 1c). 
Using this approach, the superposition is realized by preparing the ancillary atom in the superposition $\frac{1}{\sqrt{2}}(\ket{g'}+\ket{r'})$, while the measurement correspond to ancila measurement in the appropriate basis.
While these considerations are strictly valid in the case of infinite arrays, as discussed below, in practice control over atom arrays of modest size is sufficient to produce relatively large cat-like states with high fidelity (Fig. 1d). At the same time,  by appropriate choice of levels $\ket{r},\ket{r'}$, Rydberg blockade can be enabled  
over $>30\mu$m length scales  \cite{barredo2014demonstration,browaeys2016experimental}.  

Specifically, we consider the collective effect of the atoms, as their proximity allows dominant dipole-dipole interactions for the $\ket{g} \rightarrow \ket{e}$ transition. The reflection coefficient arising from the collective response of the array is given by:

\begin{align}\label{effpercollecitveEIT}
& r = \frac{i(\gamma+\Gamma)\left(\delta_r+V\right)}{ 
 {-i\left(\delta_r+V\right)(\gamma+\Gamma-2i(\delta-\Delta))+2|\Omega_p|^2}}
\end{align}
where $\Delta,\Gamma$ are cooperative corrections to the detuning ($\delta$) and decay rate ($\gamma$) associated with individual atoms 
(see SI for details). $\delta_r$  is the detuning from the Rydberg state $\ket{r}$, and $\Omega_p$ is the Rabi frequency of the control field tuned close to the $\ket{e} \rightarrow \ket{r}$ transition. The above expression accounts both for the collective effects arising from the dipole-dipole interactions within the atom array, as well as for the shift of the Rydberg level induced by the ancilla ($V$).

Let us consider a situation where the incident field is tuned to the cooperative ressonance, i.e. $\delta=\Delta$, the frequency of the pump field is chosen accordingly to match the two photon resonance ($\delta_r=0$). If the ancilla is in $\ket{g'}$ resulting in $V = 0$ for all atoms, then Eq. (\ref{effpercollecitveEIT}) fulfills the EIT condition and the array is in its transparent state $\ket{U}$, realizing $r= 0$. If the ancilla is in $\ket{r'}$ it induces a shift of the Rydberg level $V$ that can eliminate the EIT. Then $r \xrightarrow{V \rightarrow \infty} - \frac{(\gamma+\Gamma)/2}{(\gamma+\Gamma)/2-i(\delta-\Delta)}$, which on resonance results in the collective mirror state $\ket{C}$, realizing $r \rightarrow -1$. Specifically, when $V \gg \frac{|\Omega_p|^2}{(\gamma+\Gamma)/2}$ for all atoms, namely larger than the EIT window of the collective resonance, the reflection coefficient is:
\begin{equation}\label{correction2main}
r \rightarrow  -1+i\frac{|\Omega_p|^2}{(\gamma+\Gamma)/2} \frac{1}{V}.
\end{equation}
Thus, the correction from the two photon process is negligible and the collective state of the array is highly reflective ($\ket{C}$). Thus, the reflection coefficient in Eq. (\ref{effpercollecitveEIT}) is conditioned on the ancilla state. It is important to note that $V$ is inhomogeneous within the array, however, for the specific collective states we are interested in, it is sufficient to fulfill the above equality for the minimal $V$.

Before proceeding we note that the present method is a generalization of corresponding 
techniques realized  with single atoms in cavities or waveguides \cite{duan2004scalable,ritter2012elementary}. By allowing to combine intrinsic atom-photon entanglement mechanism with 
powerful techniques for shaping light propagation in free space, we now show how quantum states of multi-mode light fields can be controlled using quantum metasurfaces.

\begin{figure*}
\centering
{\includegraphics[scale = 0.35]{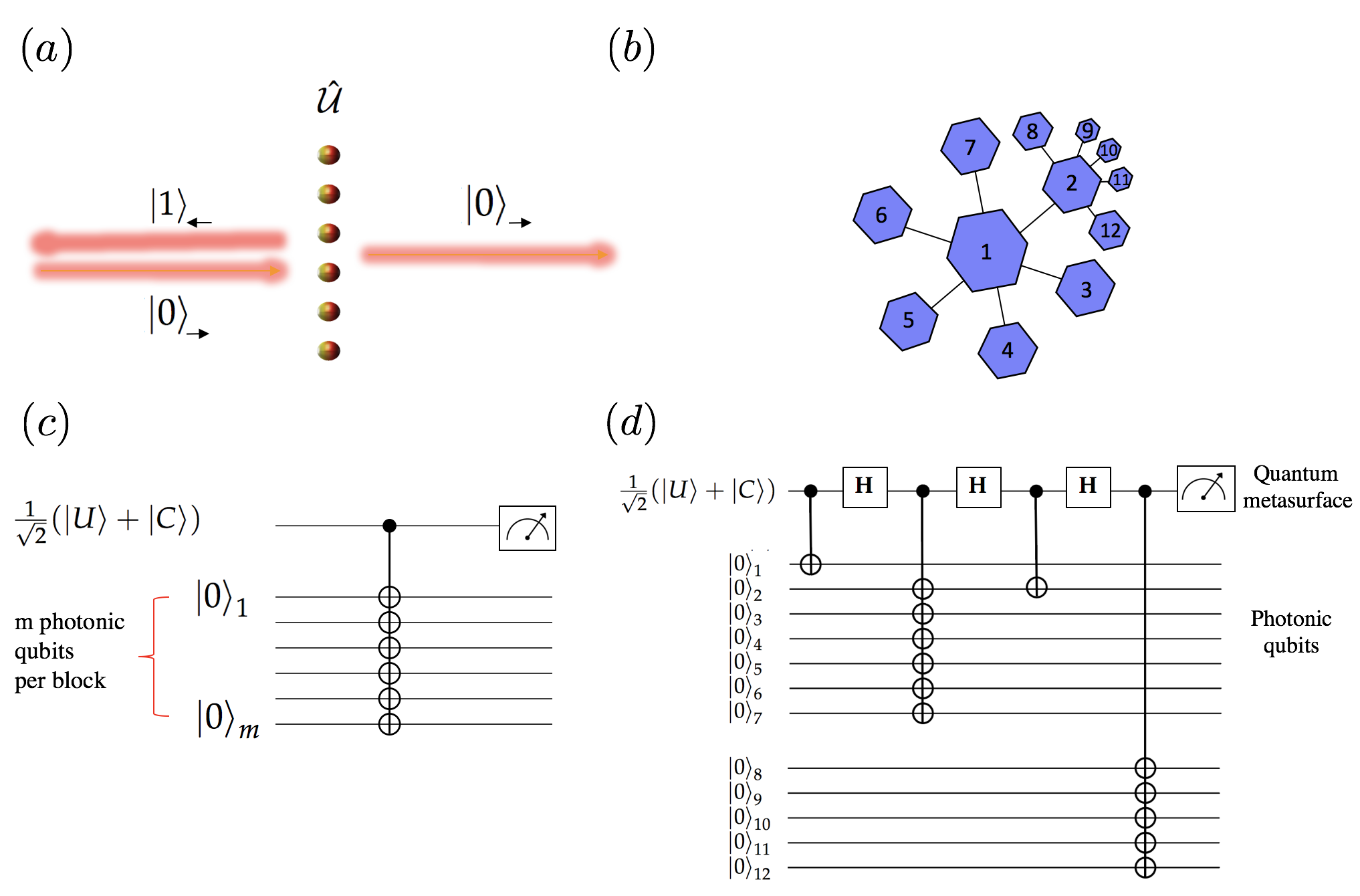}}
\caption{\textbf{Quantum information with quantum metasurfaces.} (a) Quantum optical scenario in which the quantum metasurface realizes controlled multiqubit gates on photonic qubits which are defined as right ($\ket{0}$) or left ($\ket{1}$) propagating, by reflecting or transmitting photons with defined transverse momentum. (b) Graph representation of a highly entangled tree cluster state, where the numbers indicate different photonic qubits in agreement with the numbers in (d). (c)-(d) Quantum circuits that describe preparation protocols of photonic entangled states. In both protocols the quantum metasurface is initially prepared in the state: $\frac{1}{\sqrt{2}}(\ket{U}+\ket{C})$ and at each step a parallel CNOT is applied on $m$ photonic qubits. (c) Protocol for preparing a photonic GHZ state. Following the parallel CNOT gate, a projection measurements of the quantum metasurface projects the system onto $m$ qubit photonic GHZ state. (d) Protocol for preparing the tree cluster state in (b). The procedure includes rotating the ancilla between CNOT gates and re-scattering of specific qubits to prepare the desired photonic correlations.}
\end{figure*}

\noindent
{\bf Photon state control via quantum metasurface} 

To illustrate the key idea of photon state control we first consider photonic qubits encoded in the propagation direction of single photons - either right ($\ket{0}$) or left ($\ket{1}$). The free space allows us to utilize multiple qubits by exploiting the different transverse modes.  In particular, for light propagating in the $z$ direction the natural modes can be labeled by the transverse momentum ($\textbf{k}_{\perp}$) and are defined by the electric field modes $\textbf{E}(\textbf{r},z) = \int \textbf{E}_{\textbf{k}_{\perp}}e^{\pm ikz+i\textbf{k}_{\perp} \textbf{r}} d\textbf{k}_{\perp}$, where $\textbf{k}_{\perp}$ is within the diffraction limit. 

Parallel quantum gate operations between the collective states of the atom array and photonic qubits can be employed in preparation protocols of highly entangled photonic states, which are sources for quantum communication and quantum computation \cite{knill2001scheme,raussendorf2003measurement,kuzmich2003generation,kimble2008quantum}. In these protocols, photonic qubits entangle with the quantum metasurface, which is then measured in a state that projects the system onto a desired photonic entangled state. 

One example for information processing is a parallel CNOT gate. Here the photons that encode the quantum information are defined by specific transverse momentum ($m$).
We initially prepare the quantum metasurface in the state: $\ket{\Psi_{QMS}}=\frac{1}{\sqrt{2}}(\ket{U}+\ket{C})$. A parallel CNOT gate is realized between the quantum metasurface and $m$ photonic qubits, each with defined transverse momentum ($\textbf{k}_{\perp}$), by scattering from the array. This is followed by a projective measurement of the quantum metasurface in the state: $\frac{1}{\sqrt{2}}(\ket{U}+\ket{C})$ which projects the system to a photonic GHZ state. (see Fig. 2c).
This parallel gate can serve in preparation protocols of matrix product states, which involve scattering of photons in a sequential process, that can be employed with the quantum metasurface.

\begin{figure*}
\centering
\includegraphics[scale = 0.35]{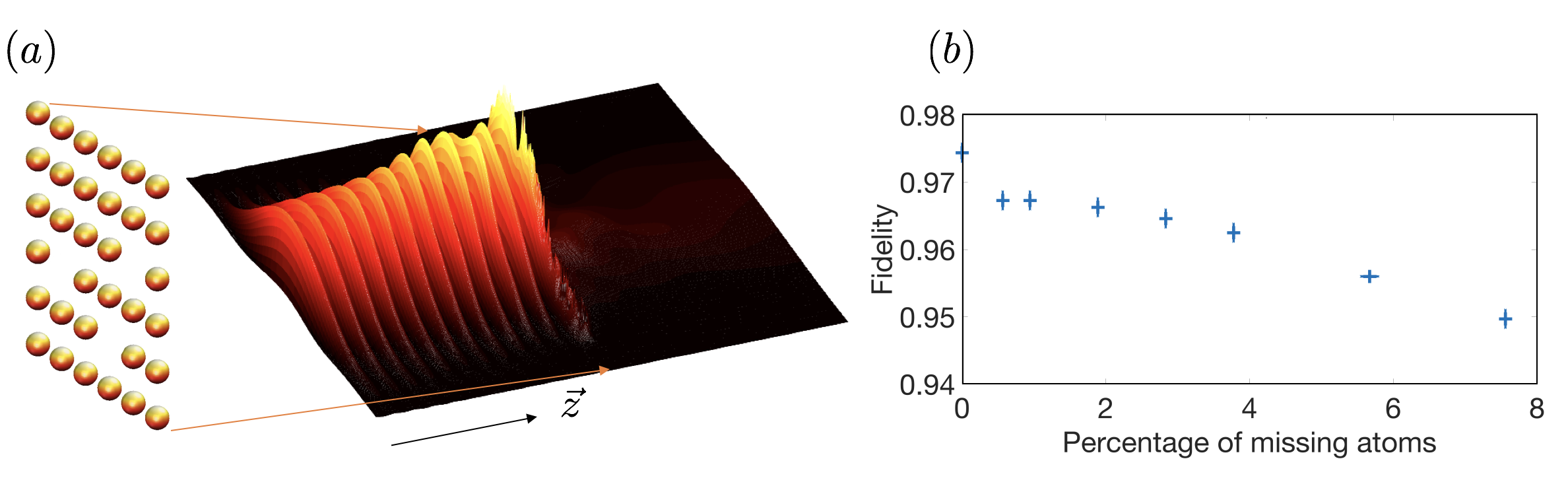}
\caption{\textbf{Effect of errors} (a) The effect of errors in the initial prepared state of the quantum metasurface on the final light state. Left: Schematic of array with an example error of missing atoms. Right: Calculated electric field distribution for scattering from an array of $25^2$ atoms with $10\%$  of the atoms missing. (b) Fidelity of the final cat light state after measuring the atom array state, as a function of the percentage of missing atoms. The fidelity is calculated for an array of $23^2$ atoms $(|\alpha|^2=9)$ and is averaged over the required number of random realizations of missing atoms, for convergence.}
\end{figure*}

As a second example, we consider preparing strong correlations of photonic tensor network states, and in particular for preparing a tree cluster state (Fig. 2b,d). The graph representation (Fig 2b) presents a tree cluster states where specific single qubits are each connected to six others. These types of entangled states are important for overcoming photon loss errors \cite{varnava2006loss}, as part of quantum error correction schemes and can serve as a basic building block to a more general entangled state which is related to holographic high energy theories \cite{pastawski2015holographic}. 
The protocol starts by initializing the quantum metasurface in the $\frac{1}{\sqrt{2}}(\ket{U}+\ket{C})$ state. After scattering a single photon, a Hadamard gate on the metasurface is applied (e.g. by driving the ancilla qubit with a $\pi$-pulse) and six new photons are launched to scatter from the quantum metasurface. Following a Hadamard gate on the ancilla, the second photon is re-scattered from the quantum metasurface for creating the desired correlations. This is followed by another Hadamard on the ancilla and scattering of five new photons from the quantum metasurface, which become entangled with the second photonic qubit. This procedure, consists of re-scattering specific photons from the quantum metasurface which can be implemented by spatial light modulators that are placed on both sides of the array to control the reflection of specific transverse modes, that need to be re-scattered (see Supplementary Information). 

Importantly, for any given protocol, the quantum information can be imprinted on any other property of the photonic qubits (flying qubits) after the scattering, and relocalized to the same optical path for further processing, by conventional linear optics (see Supplementary Information). 

The above protocols exploit the collective coupled and uncoupled states of the quantum metasurface. Since the state of the quantum metasurface controls the reflectivity of the photons in different modes, it allows realization of controlled multiqubit gate of the form:
\begin{align}\label{operator}
\hat{\mathcal{U}}= \sum_{\Phi} \ket{\Phi}\bra{\Phi} \bigotimes_{\textbf{k}_{\perp}}\mathcal{V}_{\textbf{k}_{\perp},\Phi},
\end{align}
where $\mathcal{V}_{\textbf{k}_{\perp}}$ is a unitary that depends on the scattering properties of the transverse mode $\textbf{k}_\perp$ from a quantum metasurface in state $\ket{\Phi}$. In the Supplementary Information we describe how to prepare a collective state which is reflective to specific transverse modes and transparent to others. This enables quantum control over specific photonic qubits, by preparing superpositions of mirror and non-mirror states for the different transverse modes. Analogous to a classical spatial light modulator (SLM) that gives a specific phase or amplitude to specific transverse modes, the quantum metasurface controls quantum gate operations for specific transverse modes, enabling manipulation of both spatial structure and quantum properties of the photonic state.

\noindent
{\bf Implementation, imperfections, and decoherence} 

We now discuss a specific realization of a quantum metasurface by an array of trapped neutral atoms \cite{barredo2016atom,endres2016atom,kim2016situ} whose many-body state can be manipulated by exploiting Rydberg interactions (see Supplementary Information) \cite{bernien2017probing,lienhard2018observing}. 

To realize the collective effect of the quantum metasurface atoms have to be trapped at distances that are smaller than the wavelength of the incident light. For trapping with optical tweezers the minimal spacing between atoms is above the diffraction limit. However, this requirement can be met by using, for example, $^{171}$Yb which has a telecom transition of $1389$nm and was recently trapped in optical tweezers using a laser of $470$nm \cite{covey2018telecom}. This enables the following realization of the electronic levels: $\ket{g} \equiv {6s6p}$ $ ^{ \, 3} P_{0}$ and
$\ket{e} \equiv{5d6s}$ $^{3} D_{1}$. 

The photonic state fidelity is directly affected by the atom array quality. For the EIT implementation, we estimate the error in the reflection coefficient $r$, given by Eq. (\ref{effpercollecitveEIT}).
The latter grows inversely with the shift of the Rydberg level induced by the ancilla, which is not uniform within the array (which is $\mu m$ scale). For the photonic cat state generation the resulting fidelity is 
$|\braket{\psi_{f_0}|\psi_{f}}|^2 \rightarrow 1- \frac{1}{8} |r+1|^2 |\alpha|^2$,  where $\ket{\psi_{f_0}}$ is the perfect photonic cat state. Imperfect projective measurement and the decoherence of the ancilla also affect the fidelity. 
Errors associated with imperfect preparation of atomic states or missing atoms (resulting from unsuccessful trapping) result in different scattering properties. $|\braket{\psi_{f_0}|\psi_{f}}|^2$ for different percentages of missing atoms is presented in Fig. 3b. For less than $2\%$ of missing atoms, which is a typical error in the experimental systems \cite{bernien2017probing,cooper2018alkaline}, the light state fidelity is still higher than $0.965$. 

\noindent
{\bf Outlook} 

The above considerations can be extended by combining techniques of classical metasurfaces for controlling the light degrees of freedom, such as angular momentum or polarization \cite{yu2012broadband}.  
Our analysis can be further extended to interactions between photons beyond the weak field limit \cite{lang2018interaction,lang2018non}. This work can also serve as the basis for quantum metamaterials and quantum transformation optics. Similar to their classical counterpart \cite{kildishev2013planar,kildishev2011transformation} these can be realized with systems of different dimensionality. Finally, we note that other possible experimental realizations including excitons in atomically thin semiconductors, such as transition metal dichalcogenides \cite{zhou2017probing, scuri2017atomically, scuri2018large} can also be explored to realize quantum metasurfaces.

Quantum metamaterials and metasurfaces also offer an intriguing possibility for studying analogue systems for quantum gravity. Classically, the propagation of light in a dielectric maps mathematically to motion on a space-time background, thus engineering a specific permittivity distribution which creates an analogue system for curved space  \cite{landau1971classical,genov2009mimicking,narimanov2009optical,fernandez2016anisotropic,sheng2013trapping}. The present analysis shows how one can obtain quantum control of the background itself. Thus quantum metasurfaces described here can be viewed as being analogous to a gravitational background that is itself in a quantum state. While no full theory of quantum gravity is yet fully developed, quantum metamaterials offer a novel framework to study analogues to explore consequences of possible quantum gravitational effects, such as quantum fluctuations of the gravitational background \cite{wheeler1957nature,barr2005quantum} or superpositions of background metrics \cite{zych2017bell} in a table-top system.

{\bf Acknowledgments.}
We thank Dikla Oren for helpful discussions. This work was supported through the National Science Foundation (NSF), the Center for Ultracold Atoms, the Air Force Office of Scientific Research via the Multidisciplinary University Research Initiative, and the Vannevar Bush Faculty Fellowship. R.B., I.P. and H.P. are supported by the NSF through a grant for the Institute for Theoretical Atomic, Molecular, and Optical Physics at Harvard University and the Smithsonian Astrophysical Observatory. I.P. also acknowledges funding by the Society in Science, The Branco Weiss Fellowship, administered by the ETH Zurich.

\bibliography{QuantumMetasurfaceManuscript.bib}
\nocite{*}

\end{document}



\begin{center}
  \Large\bfseries\boldmath
  Supplementary Information
\end{center}

\begin{center}
R. Bekenstein$^{1,2}$, I. Pikovski$^{1,2,3}$, H. Pichler$^{1,2}$, E. Shahmoon$^{2}$, S. F. Yelin$^{2,4}$, M. D. Lukin$^{2}$
\end{center}
\begin{center}
\textit{
$^1$ ITAMP, Harvard-Smithsonian Center for Astrophysics, Cambridge, MA 02138  \\ $^2$  Physics Department, Harvard University, Cambridge, MA 02138 \\ $^3$ Department of Physics, Stevens Institute of Technology, Hoboken, NJ 07030 \\ $^4$ Department of Physics, University of Connecticut, Storrs, CT 06269}
\end{center}

\section{Superposition of reflectivity}
The unitary transformation of a beam splitter for the two mode coherent state is:
\setcounter{equation}{0}
\renewcommand{\theequation}{S\arabic{equation}}
\begin{equation}\label{BS unitary}
B =   \begin{bmatrix}  
t & r \\ r & t 
\end{bmatrix}
\end{equation}

Where the first and second mode of the coherent state are right and left propagating modes: $\ket{\alpha_{R}, \alpha_{L}}$.  The operator realized by scattering from the quantum metasurface: 
\begin{equation}\label{quantum BS unitary}
\hat{B} = \frac{1}{\sqrt{2}} \begin{bmatrix}  
t_u & r_u \\ r_u & t_u  
\end{bmatrix} \otimes  \ket{U} \bra{U}
 +
  \frac{1}{\sqrt{2}} \begin{bmatrix}  
t_c & r_c \\ r_c & t_c  
\end{bmatrix} \otimes \ket{C} \bra{C},
\end{equation}
where $r_c,r_u$ ,$t_c,t_u$, are the reflection and transmission coefficient for the coupled state, and uncoupled state respectively. 
To prepare cat states for the light the quantum metasurface is prepared in the state: $\ket{\psi_{QMS}} = \frac{1}{\sqrt{2}}(\ket{U}+\ket{C}$). Scattering from the array results in the entangled state:
\begin{equation}\label{light state}
 \begin{aligned}
\\& \ket{\Psi} = \frac{1}{{2}} \begin{bmatrix}  
t_u & r_u \\ r_u & t_u  
\end{bmatrix}  \ket{ \begin{bmatrix}  \alpha_{R} \\  \alpha_{L} \end{bmatrix}}  \otimes \ket{U} 
 +
\frac{1}{{2}}\begin{bmatrix}   
t_c & r_c \\ r_c & t_c  
\end{bmatrix} \ket{ \begin{bmatrix}  \alpha_{R} \\  \alpha_{L}  \end{bmatrix}} \otimes \ket{C}
\end{aligned}
\end{equation}

Projection measurement of the quantum metasurface in the basis: $\ket{U} \pm \ket{C}$, projects the system to odd and even light cat states. $\hat{\mathcal{U}}$ from Eq. (4) in the main text is a generalization of Eq. (\ref{quantum BS unitary}) for a general state of the quantum metasurface, where $\mathcal{V}$ depends on the specific scattering parameters.

\section{Fidelity calculation for light state}

We calculate the fidelity of the final multiple mode coherent state of the light, in which the system is projected to after measuring the atom array state (assuming successful projection). An initial coherent state:

\begin{equation}
\ket{\psi_i} = \ket{\alpha,0}
 \end{equation}
 
is launched in perpendicular to the array and scatter from the coupled state.  The scattering results in the final state:  $\ket{t \alpha , r \alpha, \psi_{sc}}$, where $r$,$t$ are complex numbers that relate the reflected and transmitted beam to the incident beam, and $\psi_{sc}$ describes scattering to other modes. The amplitude and phase of the reflection coefficient are extracted from the numerical calculation of the scattered field, by fitting to a gaussian beam. After measuring the atom array state: $\frac{1}{\sqrt{2}}(\ket{U} +\ket{C})$ the system is projected to a light cat state:

\begin{equation}
\ket{\psi_f} = \frac{1}{\sqrt{2}}(\ket{\alpha,0,0}+\ket{t\alpha, r \alpha,\psi_{sc}}) 
\end{equation}
 
with fidelity:
\begin{equation}
 \mathcal{F}_{light}=\frac{1}{4}|(\bra{\alpha,0,0}+\bra{t \alpha, r \alpha,\psi_{sc}}) (\ket{\alpha,0,0}+\ket{0,-\alpha,0})|^2
 \end{equation}
 
By using the expression for the overlap of coherent states: $\braket{\alpha| \beta} = e^{-\frac{1}{2} |\alpha-\beta|^2}$, we calculate the fidelity, which scales exponentially with $|\alpha|^2$.



The fidelity for different array sizes is presented in Fig. 1d in the main text, for a gaussian beam with waist of $1.56\lambda$, where $\lambda$ is the wavelength. We extract $t$ and $r$, for the coupled state, by fitting to a gaussian profile and assume that the transmission is perfect for the uncoupled state. The error bars in Fig. 1d are resulted by the accuracy of fitting to a gaussian beam. In all the fidelity calculations we assume the spacing between atoms is $0.2 \lambda$ and the decay rate from the excited level is $\gamma = 24MHz$.
For calculating the Fidelity for the case of errors of missing atoms as displayed in Fig. 3b, we assume the reflected beam maintains its gaussian spatial structure, and average over different error realizations. Each data point in Fig. 3b is an average over the number of realizations necessary for numerical convergence.
Due to imperfections in the array the light is also scattered to directions that are not perpendicular to the array, resulting in the final state $\ket{t \alpha , r \alpha, \psi_{sc}}$. 

\section{Self-consistent equation for the polarizability}

To calculate the linear response of the quantum metasurface to light tuned close to the $\ket{g} \rightarrow \ket{e}$ transition, we consider the dipole-dipole interaction between the atoms which is described by the term:
\begin{equation}\label{H_int}
\hat{H}_{dd} =  \sum_{i \neq j}^N {G}(\textbf{r}_i,\textbf{r}_j) \ket{g}_i \bra{e} \otimes \ket{e}_j\bra{g}
\end{equation} 
Where $i,j$ are the atoms indices and ${G}(\textbf{r}_i,\textbf{r}_j)$ is the dyadic green function \cite{novotny2012principles} which describes the exchange of radiation between the atoms:

\begin{equation}\label{Dyadic Green}
 \begin{aligned}
{G}(\textbf{r}_i,\textbf{r}_j) = \frac{e^{ikr}}{4\pi r}[(1+\frac{ikr}{k^2r^2}) + \frac{3-3ikr -k^2r^2}{k^2r^2} \frac{|r_d|^2}{r^2}].
\end{aligned}
\end{equation}

Eq. (\ref{H_int}) in non-Hermitian as $G$ is complex, where the real part describes the dipole-dipole interaction and the imaginary part describes the collective relaxation. Since we focus on linear response to the field, we can use the non-Hermitian Hamiltonian without the need to add quantum Langevin noises.
 $k$ is the wavenumber, $r = |\textbf{r}_i-\textbf{r}_j|$, and $r_d$ is the projection of $\textbf{r} = \textbf{r}_i-\textbf{r}_j$ on the direction of the dipole matrix element. The total electric field is the sum of the incident and the scattered field:
 \begin{equation}\label{Electric field}
{E}(\textbf{r}) =  {E}_0(\textbf{r}) + 4\pi \frac{\alpha}{\epsilon_0\lambda^2} \sum_i^N {G}(k,\textbf{r},\textbf{r}_i){E}(\textbf{r}_i).
\end{equation} 
Where $\lambda$ is the wavelength, $\alpha$ is the permittivity, and $\epsilon_0$ is the vacuum permittivity.
As done in \cite{shahmoon2017cooperative} from Eq. (\ref{Electric field}) we find the self-consistent equation for the polarizability, by calculating the electric field in the location of the atoms. Taking $\textbf{r} = \textbf{r}_j$, and recalling the linear relation between the polarizability and the electric field $ {P} = \alpha  {E}$:
 \begin{equation}\label{P}
{P}_i =  {P}_{i,0}+ \sum_{i \neq j}^N 4\pi^2 \frac{\alpha}{\epsilon_0 \lambda^2} {G}(k,\textbf{r}_i,\textbf{r}_j) {P}_j
\end{equation} 

We use Eq. (\ref{Electric field}) to calculate the scattered field from the polarizability built on the atoms, resulting in the results displayed in Fig. 1d.

\section{Electromagnetically induced transparency for the collective resonance}

We hereby calculate the effective permittivity for light tuned to the $\ket{g} \rightarrow \ket{e}$ transition for the cascade three level system in Fig 1b. The non-hermitian Hamiltonian of the array is: 

\begin{align}\label{coherencesge2}
 \begin{aligned}
& \hat{H} =  \sum_{i}^N [ -\hbar \omega_e \ket{e}_i \bra{e}_i  \\ &  -  (\ket{g}_i \bra {e}_i \mu_1 + \ket{e}_i \bra{g}_i \mu_1)({\epsilon_1} e^{-i \nu_1 t} +{\epsilon_1}^* e^{i \nu_1 t}) \\ & - (\ket{e} _i \bra {r}_i \mu_2 + \ket{r}_i \bra{e}_i \mu_2)({\epsilon_2} e^{-i \nu_2 t}+{\epsilon_2}^* e^{i \nu_2 t})] \\ &  -\hbar \omega_r \ket{r}_i \bra{r}_i 
+  \frac{3\pi \gamma c \hbar}{\omega_e}\sum_{i<j}^N G(\textbf{r}_i,\textbf{r}_j) \ket{e}_i\bra{g}_i \otimes \ket{g} _j \bra{e}_j,\end{aligned}
\end{align} 

where $\epsilon_{1,2}$ is an external electric field and $\mu_{1,2}$ is the dipole matrix element relevant to the dipole allowed transitions. In the last term $G$ is the dyadic green function.  In the situation where the system starts in the ground state and the probe field (tuned to the $\ket{g} \rightarrow \ket{e}$ resonance) is weak, we can treat the system by adding terms to incorporate decay channels from $\ket{r}$ and $\ket{e}$ defining $\gamma_{r}$ and $\gamma$: $\hat{H}_{eff} = \hat{H} -  \sum_{i}^N \frac{i\hbar \gamma}{2} \ket{e}_i\bra{e}_i - \frac{i\hbar \gamma_r}{2} \ket{r}_i\bra{r}_i$. 

For a translation invariant array the dyadic green function - $G(\textbf{r}_i,\textbf{r}_j)$ can be diagonalize in momentum space: $G_{\textbf{k}} = \sum_{i \neq j} e^{i \textbf{k}_{\perp} (\textbf{r}_i-\textbf{r}_j)} G(\textbf{r}_i,\textbf{r}_j)$. We transform the Hamiltonian to the basis of collective excitations with defined momentum:
 
 \begin{align}\label{coherencesge2}
 \begin{aligned}
&  \hat{H}_{eff} =  \sum_{\textbf{k}} -\hbar \omega_e \ket{e} _{\textbf{k}} \bra{e}_{\textbf{k}} \\ &  -  (\ket{g}_{\textbf{k}} \bra {e}_{\textbf{k}} \mu_1 + \ket{e}_{\textbf{k}} \bra{g}_{\textbf{k}} \mu_1)({\epsilon_{1,\textbf{k}}} e^{-i \nu_1 t} +{\epsilon_{1,\textbf{-k}}}^* e^{i \nu_1 t})  \\ & +   3\gamma \lambda \hbar G_{\textbf{k}} \ket{e}_{\textbf{k}} \bra{e}_{\textbf{k}} -\hbar \omega_r \ket{r} _{\textbf{k}}  \bra{r}_{\textbf{k}}   \\ &- (\ket{e}_{\textbf{k}} \bra {r}_{\textbf{k}} \mu_2 + \ket{r} _{\textbf{k}} \bra{e}_{\textbf{k}} \mu_2)({\epsilon_{2}} e^{-i \nu_2}+{\epsilon_{2}}^* e^{i \nu_2 t})  \\ &  - i\hbar \frac {\gamma}{2} \ket{e}_{\textbf{k}} \bra{e}_{\textbf{k}} - i \hbar \frac{ \gamma_{r}}{2} \ket{r}_{\textbf{k}} \bra{r}_{\textbf{k}}
\end{aligned}
\end{align} 

Where $\ket{e}_{\textbf{k}} = \frac{1}{\sqrt{N}} \sum_i^N e^{ i \textbf{k}_{\perp} \textbf{r}_i} {\sigma_{eg,i}}^{+} {\ket{g}}^{\otimes N}$ (with ${\sigma_{eg,i}}^{+} = \ket{e}_i\bra{g}_i$), and
$\ket{r}_{\textbf{k}} =  \frac{1}{\sqrt{N}} \sum_i^N {\sigma_{re,i}}^{+} e^{ i \textbf{k}_{\perp} \textbf{r}_i} {\sigma_{eg,i}}^{+} {\ket{g}}^{\otimes N}$ (with ${\sigma_{re,i}}^{+} = \ket{r}_i\bra{e}_i$), and $\epsilon_{\textbf{k}} =\sum_i \epsilon_1e^{-\textbf{k}_{\perp}\textbf{r}_i}$, where $\textbf{k}_{\perp}$ is the transverse momentum as defined before. We define the wavefunction as: $\ket{\psi} = c_{g}(t) \ket{g}^{\otimes N} + c_{e,\textbf{k}}(t) \ket{e}_{\textbf{k}} + c_{r,\textbf{k}}(t) \ket{r}_{\textbf{k}}$. We assume the probe field is weak and the excited level $\ket{e}_\textbf{k}$, and Rydberg level $\ket{r}_\textbf{k}$ populations are small and solve the Schrodinger equation (assuming small quantum jumps). 
We then transform to the rotating frame: $\tilde{c}_{e,\textbf{k}}(t) = {c_{e,\textbf{k}}(t)} e^ {-i \nu_1 t}$, $\tilde{c}_{r,\textbf{k}}(t) = {c_{r,\textbf{k}}(t)} e^ { i(\nu_2 t-\nu_1 t)}$, writing the coupled equations for the coefficients $c_{g}(t), c_{e,\textbf{k}}(t), c_{r,\textbf{k}}(t)$:

\begin{equation}\label{s1}
\dot{c}_{g} = i {{\Omega_{\textbf{k}}} ^*} c_{e,\textbf{k}}  
\end{equation} 

\begin{align}\label{s2}
\dot{c}_{r,\textbf{k}} = -(\gamma_{r}/2 - (\delta_{r} +V))+ {\Omega_{p}^*} c_{e,\textbf{k}}  
\end{align} 

\begin{align}\label{s3}
\dot{c}_{e,\textbf{k}} = -(\gamma+\Gamma_{\textbf{k}})/2 - i (\delta-\Delta_{\textbf{k}}))+ i{\Omega_{\textbf{k}}}  c_{g} +{\Omega_p} c_{r,\textbf{k}}
\end{align} 

The detuning and the decay rates of the collective excitations has corrections relative to the bare atom permittivity. Specifically, these depend on the momentum:
$\Delta_{\textbf{k}} = -3 \gamma \lambda \Re(G_{\textbf{k}})$ and $\Gamma_{\textbf{k}} =\frac{3}{2} \gamma \lambda \Im(G_{\textbf{k}})$.
Where the Rabi frequencies are: $\Omega_p = \frac{\epsilon_{2} \mu_2}{\hbar} $ and $\Omega_{\textbf{k}} = \frac{\epsilon_{1,\textbf{k}} \mu_1}{\hbar} $, and the detunings are: $\delta = \nu_1 - \omega_e$ and $\delta_{r}= ( \nu_1-\omega_{e})-(\nu_2-\omega_{r})$.

$V$ is caused by Rydberg interaction induced by the ancilla. It is important to note that $V$ is inhomogeneous within the array. However for the specific collective states in interest, it is sufficient to assume $V$ in homogeneous.

We solve equations Eqs. (\ref{s1})-(\ref{s3}) for a weak field probe assuming the population is mostly in the ground state, and derive the coherence: ${\rho}_{eg,\textbf{k}} = {c_g}^{*} c_{e,\textbf{k}}$:
\begin{equation}\label{coherencesge1}
{\rho}_{eg,\textbf{k}} =  \frac{i \Omega_{\textbf{k}}(\gamma_{r}/2-i(\delta_{r}+V))}{\left((\frac{\gamma_r}{2}-i(\delta_{r}+V))(\frac{\gamma+\Gamma_{\textbf{k}}}{2}-i(\delta-\Delta_{\textbf{k}}))\right)+|\Omega_p|^2}
\end{equation} 


The effective permittivity is propotional to the coherence: $ \alpha({\textbf{k}}) = \frac{|\mu|^2}{\hbar} \frac{\rho_{eg,\textbf{k}}}{\Omega_{\textbf{k}}} $:

\begin{equation}\label{eff_EIT}
\alpha_{eff}({\textbf{k}}_{\perp})=  \frac{i|\mu|^2(\gamma_{r}/2-i(\delta_{r}+V))/\hbar}{\left((\frac{\gamma_r}{2}-i(\delta_{r}+V))(\frac{\gamma+\Gamma_{\textbf{k}}}{2}-i(\delta-\Delta_{\textbf{k}}))\right)+|\Omega_p|^2}
\end{equation}

We rewrite Eq. (\ref{eff_EIT}) in order to see the effect of the two photon resonance:
\begin{align}\label{effpercollecitveEITsupp}
 \begin{aligned}
& \alpha_{eff}({\textbf{k}}_{\perp}) = \frac{|\mu|^2}{\hbar} \frac{i}{(\gamma+\Gamma_{\textbf{k}})/2-i(\delta-\Delta_{\textbf{k}})} \times  (1- \frac{\frac{|\Omega_p|^2}{(\gamma+\Gamma_{\textbf{k}})/2-i(\delta-\Delta_{\textbf{k}})}}{\gamma_{r}/2-i(\delta_{r}+V)+\frac{|\Omega_p|^2}{((\gamma+\Gamma_{\textbf{k}})/2-i(\delta-\Delta_{\textbf{k}}))}}
 \end{aligned}
\end{align} 

The first term in Eq. (\ref{effpercollecitveEITsupp}) is the collective single photon resonance (which gives rise to the perfect mirror effect for the right detuning) and the second is the two photon resonance, which is affected by the shift of the Rydberg level.  

We are interested in the reflection coefficient for the collective states $\ket{U}$ and $\ket{C}$.  For both cases we assume the probe light is tuned to the collective resonance: $\delta-\Delta_{\textbf{k}} = 0$, and that $\delta_r = 0$ and $\gamma_r = 0$.

If the ancilla is in the $\ket{g'}$ state, then $V=0$ and $\ket{U}$ is realized as the EIT condition is fulfilled. In this case ${\alpha_{eff}} = 0$, resulting in a transparent array (with reflection coefficient $r=0$).

If the ancilla is in the $\ket{g'}$ state, realizing $V \gg \frac{|\Omega_p|^2}{(\gamma+\Gamma_{\textbf{k}})/2}$ for all atoms, then $\ket{C}$ is realized. This results in a collective permittivity that agrees with \cite{shahmoon2017cooperative} with the correction: 
\begin{equation}\label{correction2}
\Delta \alpha_{eff}(\textbf{k}_{\perp}) \rightarrow \frac{|\mu|^2}{\hbar} \frac{|\Omega_p|^2}{({\gamma}+\Gamma_{\textbf{k}})^2} \frac{1}{V}
\end{equation} 

For $ \frac{|\mu|^2}{\hbar} \frac{|\Omega_p|^2}{({\gamma}+\Gamma_{\textbf{k}})^2} \ll V$ this term is negligible and the collective  resonance gives a mirror effect.   
We rewrite Eq. (\ref{eff_EIT}) with the spontaneous emission rate $\gamma = \frac{|\mu|^2 k^3}{3 \pi \epsilon_0 \hbar}$, and derive the reflection coefficient in (Eq. (2)) in the main text by scattering theory (presented in section 2.3 in \cite{shahmoon2017cooperative}), for the above two cases, and for light launched perpendicular to the array.




\section{Quantum information with quantum metasurfaces}

Here we present a concrete quantum optical scenario in which the quantum metasurface realizes a unitary $\hat{\mathcal{U}}$ which applies controlled multiqubit gates on photonic qubits, which are defined as right ($\ket{0}$) or left ($\ket{1}$) propagating. For relocalizing the quantum information after the scattering the transmitted photons to the right side of the quantum metasurface are transmitted through a half-waveplate that converts the handedness of their polarization. By mirrors and a polarizing beam splitter (PBS) the quantum information can be relocalized to the negative $z$ direction (See Fig. S1). 

As the quantum metasurface realizes the unitary $\hat{\mathcal{U}}$ in Eq. (4) it can be used as an ancilla to prepare highly entangled states for photons. For convenience we define: $\ket{+}_{QMS} \equiv \frac{1}{\sqrt{2}}(\ket{U}+\ket{C})$.

A one-dimensional cluster state is realized by:
\begin{equation}\label{clusterstate}
\ket{\psi_{1D}} = (\prod_{\textbf{k}_{\perp}=1}^M  \hat{H}_{QMS}  \hat{X}_{QMS,\textbf{k}_{\perp} }) \ket{+}_{QMS} \otimes \ket{0}_{\textbf{k}_{\perp}}^{\otimes M}
\end{equation} 	

\begin{figure*}
\centering
\includegraphics[scale = 0.35]{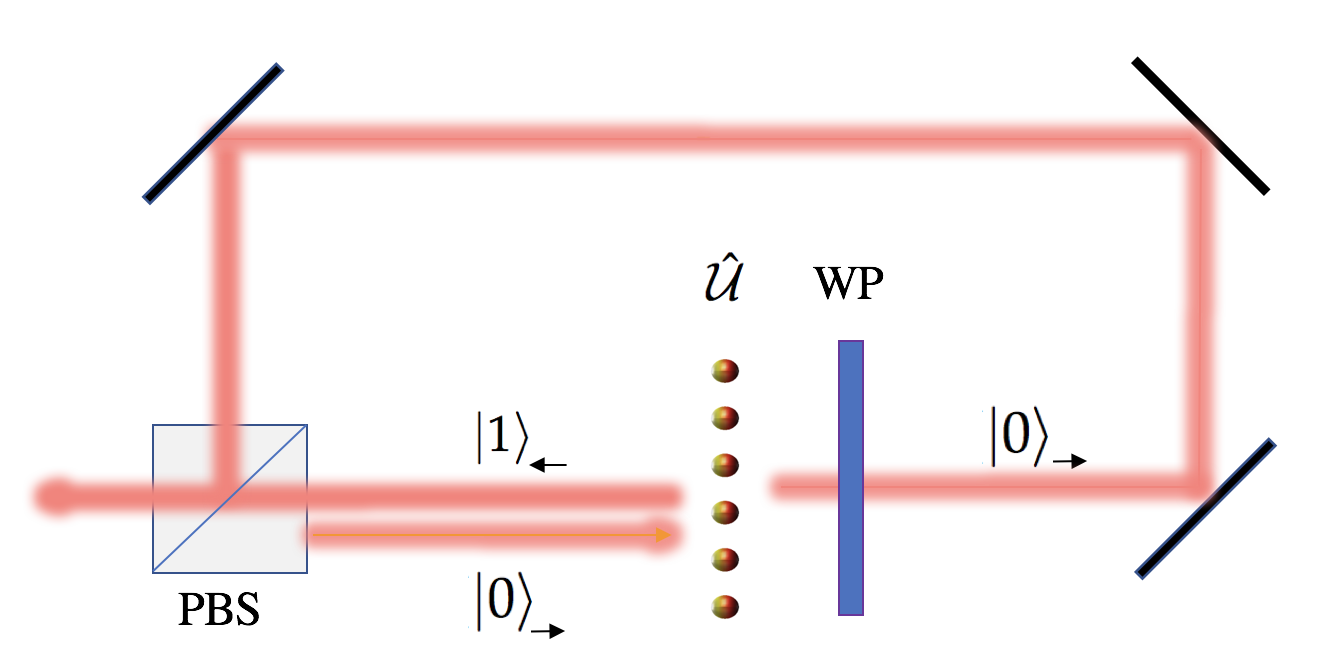}
\caption{\textbf{Quantum optical scenario with the quantum metasurface.}  The quantum metasurface apply multi qubit gates on photonic qubits by scattering. For relocalizing the quantum information after the scattering, the transmitted photons to the right side of the quantum metasurface are transmitted through a half-waveplate that converts the handedness of their polarization. After the information is encoded on the polarization, the photons are relocalize to the negative $z$ direction by mirrors and a polarizing beam splitter (PBS).}
\end{figure*}

Where $\hat{H}_{QMS}$ is a global Hadamard on the quantum metsurface, and $\hat{\tilde{X}}_{QMS,{\textbf{k}_{\perp}}}$  is a CNOT on the qubit $\textbf{k}_{\perp}$ where the quantum metsurface is the controlled qubit. We further extend the gate to a parallel CNOT gate on multiple qubits:

\begin{equation}\label{CNOT on multiple qubits}
\hat{\mathcal{X}}_{QMS,\textbf{k}_{\perp}}= \ket{C}\bra{C} \bigotimes_{\textbf{k}_{\perp}} \hat{X}_{QMS,\textbf{k}_{\perp}}+\ket{U}\bra{U} \bigotimes_{\textbf{k}_{\perp}} \mathbb{1}_{\textbf{k}_{\perp}} 
\end{equation} 

To prepare a GHZ state:
\begin{equation}\label{clusterstate2}
\ket{GHZ} =  \hat{\mathcal{X}}_{QMS,\textbf{k}_{\perp}=1,2...M} \ket{+}_{QMS} \bigotimes_{\textbf{k}_{\perp}} \ket{0}_{\textbf{k}_{\perp}} ^{\otimes M}
\end{equation} 

We then measure the quantum metasurface in the state $\ket{+}_{QMS}$ to prepare the photonic GHZ state (Fig. 2c in the main text). 

\section{Atomic excitations with defined momentum}

In order to exploit additional degrees of freedom of the quantum metasurface Hilbert space we discuss superpositions of distinct atomic state. We assume that each atom has three electronic levels (see Fig. S2a): two metastable states ($\ket{g_1},\ket{g_2}$) and an excited state $\ket{e}$ with radiative transitions to $\ket{g_1}$, while $\ket{g_2}$ is far detuned from the incident light tuned close to the $\ket{g_1} \rightarrow \ket{e}$ transition. Thus, atoms in $\ket{g_1}$ are coupled (with bare permittivity $\alpha_0$), whereas atoms in $\ket{g_2}$ are uncoupled to the radiative transition. As a simple example, by preparing the array in the macroscopic superposition of distinct atomic states: $\frac{1}{\sqrt{2N}}(\ket{g_1}^{\otimes N}+\ket{g_2}^{\otimes N})$,  the collective state $\frac{1}{\sqrt{2}} (\ket{C} + \ket{U})$ is realized.

States with spatial variation of $\ket{g_1}$ or $\ket{g_2}$ population, can control the reflectivity properties of the quantum metasurface. Spatial-structured states with crystalline order were demonstrated in \cite{bernien2017probing}, and can be described by a range of atomic momentum. 
We give an example of collective states that are reflective to specific transverse modes and transparent to others. We focus on states with one atomic wavevector $\ket{K_a}$, which realize periodic modifications of the classical bare permittivity:


\begin{figure*}
\centering
\includegraphics[scale = 0.35]{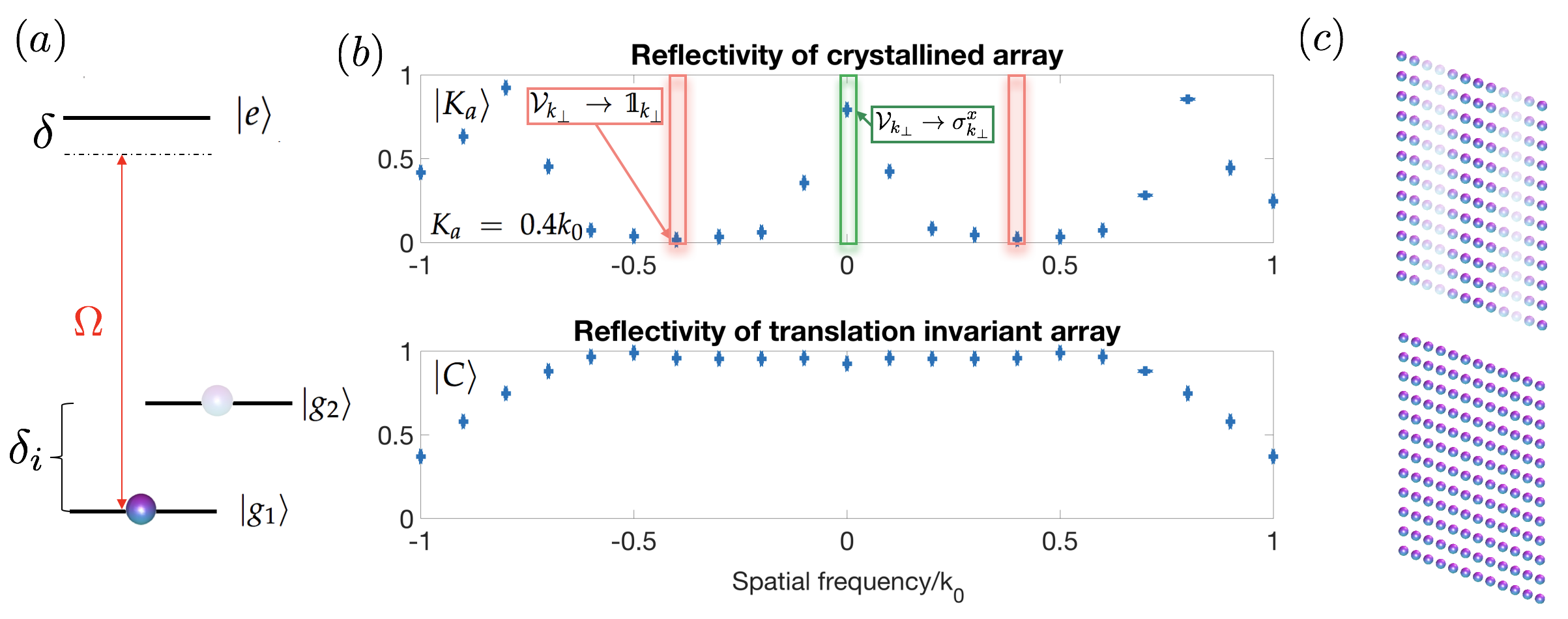}
\caption{\textbf{Controlled reflectivity of photons with specific transverse momentum.} (a) Electronic levels of the atoms in the array for quantum metasurface implementation with distinct atomic superposition, where $\ket{g_1}$ and $\ket{g_2}$ are coupled and uncoupled to the excited level, respectively. (b) Calculated reflectivity $|r|^2$ as a function of the light's transverse momentum, for atom arrays with permittivity configurations as illustrated in (c). Upper label: reflectivity of atom array with periodic perturbation of the permittivity ($K_a = 0.4k_0$) which is unity for modes in the green area but drops to $ \approx 0.01$ for modes in the red area. Relating to Eq. (4), the unitary $\hat{\mathcal{U}}$ is realized with $\mathcal{V}_{\textbf{k}_{\perp},\Phi} \rightarrow {\sigma^x_{\textbf{k}_{\perp}}}$ for the modes in the green area and $\mathcal{V}_{k_{\perp},\Phi} \rightarrow \mathbb{1}_{\textbf{k}_{\perp}}$ for modes in the red area. Lower label: reflectivity of a translation invariant array ($K_a = 0$) which is realized by the collective atomic state $\ket{C} = \frac{1}{\sqrt{N}} \ket{g_1}^{\otimes N}$. (c) Illustration of the atom permittivity $\alpha_i$, for collective states which their reflectivity is displayed in (b). The color transparency indicates $\alpha_i$ (fully transparent is $\alpha_i = 0$).}
\end{figure*}


\begin{equation}\label{permittivity}
\braket{\alpha_i} = \bra{K_a}  \alpha_i  \ket{K_a} = \alpha_0 \frac{1+cos(2 K_a \textbf{r}_i)}{2}
\end{equation} 

In the limit of infinite array the polarizability in momentum space is (Eq. (\ref{P}):
\begin{equation}\label{polarizability}
{P}(\textbf{k}_{\perp})  = {P}_0(\textbf{k}_{\perp}) + \frac{4 \pi^2}{\epsilon_0 \lambda^2} \frac{1}{N} \sum_i^N \sum_{i \neq j}  e^{-i\textbf{k}_{\perp} \textbf{r}_i} \alpha_i  {G}(k, \textbf{r}_i, \textbf{r}_j) {P}_j
\end{equation} 

Where the Fourier transform is defined on atom locations: $ P(\textbf{k}_{\perp}) = \frac{1}{N}\sum_{n=1}^N e^{i \textbf{k} _{\perp} \textbf{r}_n} P(\textbf{r}_n)$ (where $\textbf{r}_n$ is the location of the $n$ atom).
Specifically, for $\alpha_i$ defined by Eq. (\ref{permittivity}):




\begin{equation}\label{polarizability4}
{P}(\textbf{k}_{\perp})  = {P}_0(\textbf{k}_{\perp}) + \frac{4 \pi^2}{2 \epsilon_0 \lambda^2} \sum_{s = 0,+,-}     {G}(\textbf{k}_{\perp} - s K_{a}){P}(\textbf{k}_{\perp}- K_{a}),
\end{equation} 
here, ${G}$ is the Fourier transform of Green's function ${G}(k, \textbf{r}_i, \textbf{r}_j)$. Generally, the green function couples different $\textbf{k}_{\perp}$s. 
We calculate the effective permittivity (which relates the incoming field ${E}_0$ to the polarizability built on the atoms (${P}$)) in the basis of the eigen-functions, for which the dyadic green function is diagonal ($u_m$). We derive the polarizability:

\begin{equation}\label{polarizability5}
\begin{aligned}
{P}( \textbf{k}_{\perp})  \approx  \sum_m |f_{m \textbf{k}_{\perp}}|^2  \braket{u_m | \alpha_{eff} |u_m} {E}_0(\textbf{k}_{\perp}),
\end{aligned}
\end{equation}

where $f_{m \textbf{k}_{\perp}} = \sum_i {u}_m e^{-i \textbf{k}_{\perp} \textbf{r}_i} d \textbf{r}_i$. The approximate sign appears due to the neglection 
of off diagonal terms that are numerically orders of magnitude smaller than diagonal terms.
The reflectivity presented in Fig. S2 is calculated by Eq. (\ref{polarizability5}) using scattering theory \cite{shahmoon2017cooperative}, for wavevector with one component: $(K_a,0$). 
For comparison, for an array in the collective state $\frac{1}{\sqrt{N}}\ket{g_1}^{\otimes N} = \ket{C}$ the reflectivity is unity for all spatial modes within the diffraction limit (Fig. S2b lower label). However, for an array with a specific periodic modification of the permittivity, the reflectivity drops for transverse modes in a specific region in momentum space (illustrated in red in Fig. S2b upper label), realizing $\mathcal{V}_{\textbf{k}_{\perp},\Phi} \rightarrow \mathbb{1}_{\textbf{k}_{\perp}}$.  
This state of the array is transparent for specific photonic transverse momentum $\textbf{k}_{\perp} = K_a$, similar to phenomena existing in photonic crystals \cite{fan2002analysis}.
This enables quantum control over specific photonic qubits, by preparing superpositions of mirror and non-mirror states for the different transverse modes. Analogous to a classical spatial light modulator (SLM) that gives a specific phase or amplitude to specific transverse modes, the quantum metasurface controls quantum gate operations for specific transverse modes (see Eq. (4) in the main text). However, it is important to highlight the difference from the classical counterparts (as SLM and phase masks): coupling between different wavevectors and imperfection in reflectivity is a serious obstacle for quantum information applications, as oppose to classical photonic applications.

\section{Manipulating the many body state via Rydberg interactions}

We now discuss preparation of superpositions of distinct atomic states. One way to manipulate the many-body state of the atom array is by employing an ancillary atom with a ground state $\ket{g'}$ and a highly excited Rydberg state $\ket{r'}$. For preparing general many body states of the atom array, we condition the evolution of the quantum metasurface on the ancilla state. In particular, by applying local classical fields, a Raman transition of atoms from $\ket{g_1}$ to $\ket{g_2}$ through the Rydberg level (Fig. S3) gives rise to the unitary \cite{jaksch2000fast}:

\begin{equation}\label{ancilla gate}
\begin{aligned}
 \hat{\mathcal{U}_a} =  \ket{g'} \bra{g'}  \otimes \hat{\mathcal{W}}+\ket{r'} \bra{r'}  \otimes \mathbb{1},
 \end{aligned} 
\end{equation}

which can be used to prepare entangled state of the atom array. In particular if  the ancillary atom is initially prepared in $\frac{1}{\sqrt{2}}(\ket{g'}+\ket{r'})$ and all atoms in the state $\ket{g_1}$. Subsequently the atoms of the metasurface are flipped to the state $\ket{g_2}$, if the ancilla is in $\ket{g'}$, i.e. $\hat{\mathcal{W}}=\prod_i \sigma_{12,i}^x$ (where $\sigma_{12,i}^x= \ket{g_1}_i\bra{g_2}_i+\ket{g_2}_i\bra{g_1}_i$). 

 \begin{figure*}
\centering
\includegraphics[scale = 0.35]{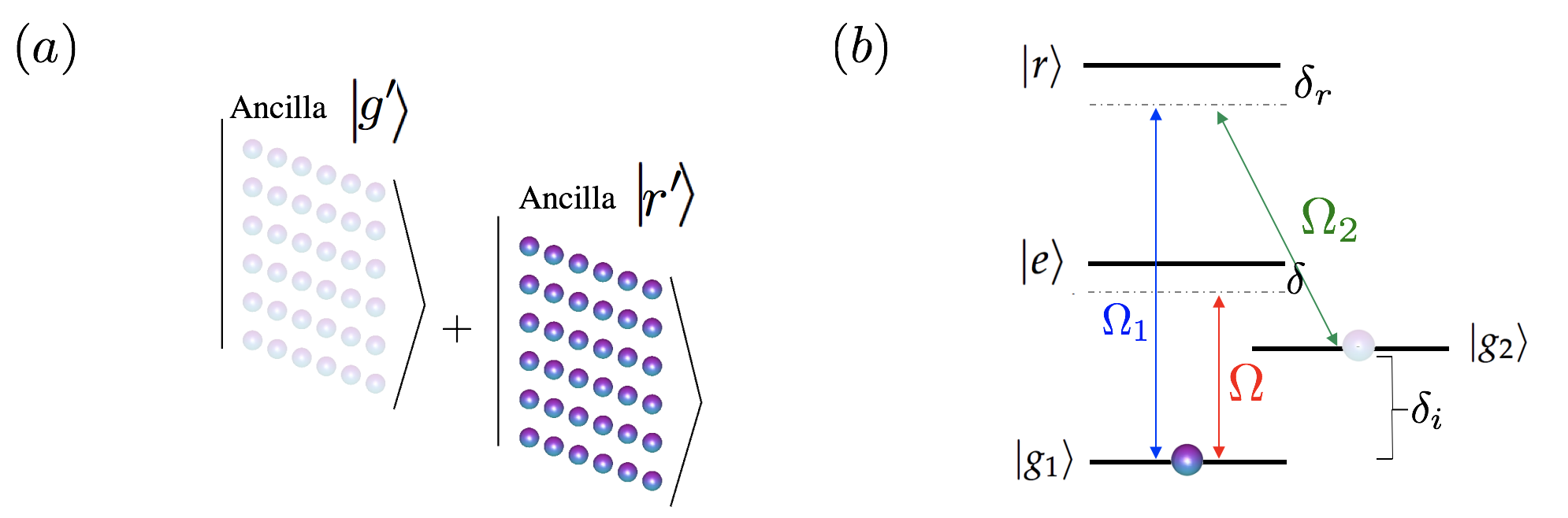}
\caption{\textbf{Realization of quantum metasurfaces with atoms excited to Rydberg states} (a) Ancillary atom prepared in a superposition of ground and Rydberg state controls the state of the array. (b) Energy levels of atoms in the array for preparation of GHZ state. Population can be coherently transferred from $\ket{g_1}$ to $\ket{g_2}$ by a Raman transition through the Rydberg level $\ket{r}$, applying the control fields $\Omega_1$ and $\Omega_2$.}
\end{figure*}

A final measurement of the ancillary atom in the basis $\frac{1}{\sqrt{2}}(\ket{g'}\pm\ket{r'})$ prepares the quantum metasurface in the desired quantum many-body state. In a similar manner the definition of $\hat{\mathcal{W}}$ can be generalized to excite atom arrays to collective states that are coupled and uncoupled to specific transverse modes as described above.

For distinct atomic superposition states the fidelity of the quantum metasurface state limits the light state fidelity also by the projective measurement. In addition, specific errors in the atom array state preparation, affect the fidelity of the photonic state directly by modifying the scattering properties. 
For simplicity, we assume the depolarization channel $\epsilon$ is uniform for all atoms, then the fidelity of the atom array state is $(1-\epsilon)^N$ resulting the photonic state fidelity: $\mathcal{F}_{light} = (1-\epsilon)^N |\braket{\psi_{f_0}|\psi_{f}}|^2$. In Fig. 3 in the main text we present $|\braket{\psi_{f_0}|\psi_{f}}|^2$ for example errors.

\bibliography{SuppQuantumMetasurfaceManuscript.bib}
\nocite{*}